\documentclass[aps,prl,floatfix,twocolumn,footinbib]{revtex4-2}

\usepackage{epsfig}
\usepackage{graphicx}
\usepackage{amsmath,amssymb,bm}
\usepackage{color}

\usepackage[sort&compress]{natbib}
\parskip=\medskipamount

\newcommand{\eq}[1]{(\ref{#1})}

\newcommand{\be}{\begin{equation}}
\newcommand{\ee}{\end{equation}}

\newcommand{\appropto}{\mathrel{\vcenter{
  \offinterlineskip\halign{\hfil$##$\cr
    \propto\cr\noalign{\kern2pt}\sim\cr\noalign{\kern-2pt}}}}}

\begin{document}

\title{Universal scaling of segment fluctuations  in polymer and chromatin dynamics}

\author{Kirill E. Polovnikov$^{1,2,\;}$} \email{kipolovnikov@gmail.com} 
\author{Mehran Kardar$^{3,\;}$} \email{kardar@mit.edu}

\affiliation{$^1$ Skolkovo Institute of Science and Technology, 121205 Moscow, Russia \\
$^2$ Institute for
Physics \& Astronomy, University of Potsdam, 14476 Potsdam-Golm, Germany \\
$^3$ Department of Physics, Massachusetts Institute of Technology, Cambridge, USA}

\begin{abstract} 
We demonstrate how center-of-mass (COM) motion influences polymer segment fluctuations. Cancellation of internal forces, together with spatially uncorrelated external noise, generally yields COM diffusivity scaling as $1/s$ with segment length $s$, regardless of fractal dimension, viscoelasticity, or activity. This introduces distinct dynamic scaling corrections to two-point fluctuations and quenched-induced tangential correlations, validated by theory, simulations, and chromatin imaging data. In the latter, the extracted dynamic exponent reveals topological constraints, thereby resolving the discrepancy between chromatin's crumpled structure and its Rouse-like dynamics.
\end{abstract}

\maketitle

\textit{Introduction:} Polymer dynamics in complex environments has been a central theme in soft matter physics since de Gennes' pioneering work on reptation in obstacle arrays~\cite{DeGennes1979} and the subsequent development of the tube model by Doi and Edwards~\cite{doi}. Classical Rouse and Zimm models describe ideal chain dynamics; however,  real polymeric systems exhibit more complex behavior due to topological constraints and memory effects~\cite{TammPolovnikov,GKh94,khokhlov85,rubinstein86,everaers04}, leading to non-ideal chain statistics, as seen in melts of unknotted non-concatenated polymer rings \cite{rosa14,halverson11_1}. 

The emergence of high-resolution experimental data on polymer systems—such as chromatin~\cite{keizer23,gabriele22,bruckner}—has motivated renewed examination of classical results in polymer dynamics~\cite{tortora20}. 
Theoretical models of chromatin often build on the Rouse framework, incorporating additional physical ingredients such as viscoelasticity~\cite{polovnikovprl18,weber10,lampo16}, fractal folding of the chromatin fiber~\cite{polovnikovprl18,tamm15,polovnikovsm18,Amitai13}, and active processes~\cite{Weber12,Goychuk23,Zidovska13,Eshghi22,Shin23,chan24}. While the resulting dynamical exponents can often be related through scaling arguments à la de Gennes~\cite{polovnikovprl18,tamm15,bruckner},  derivations from underlying physical mechanisms are frequently lacking. A deeper understanding that identifies universal dynamical features emerging from collective polymer motion, without relying on detailed model assumptions, is therefore desirable.


\begin{figure}
\includegraphics[width=0.48\textwidth]{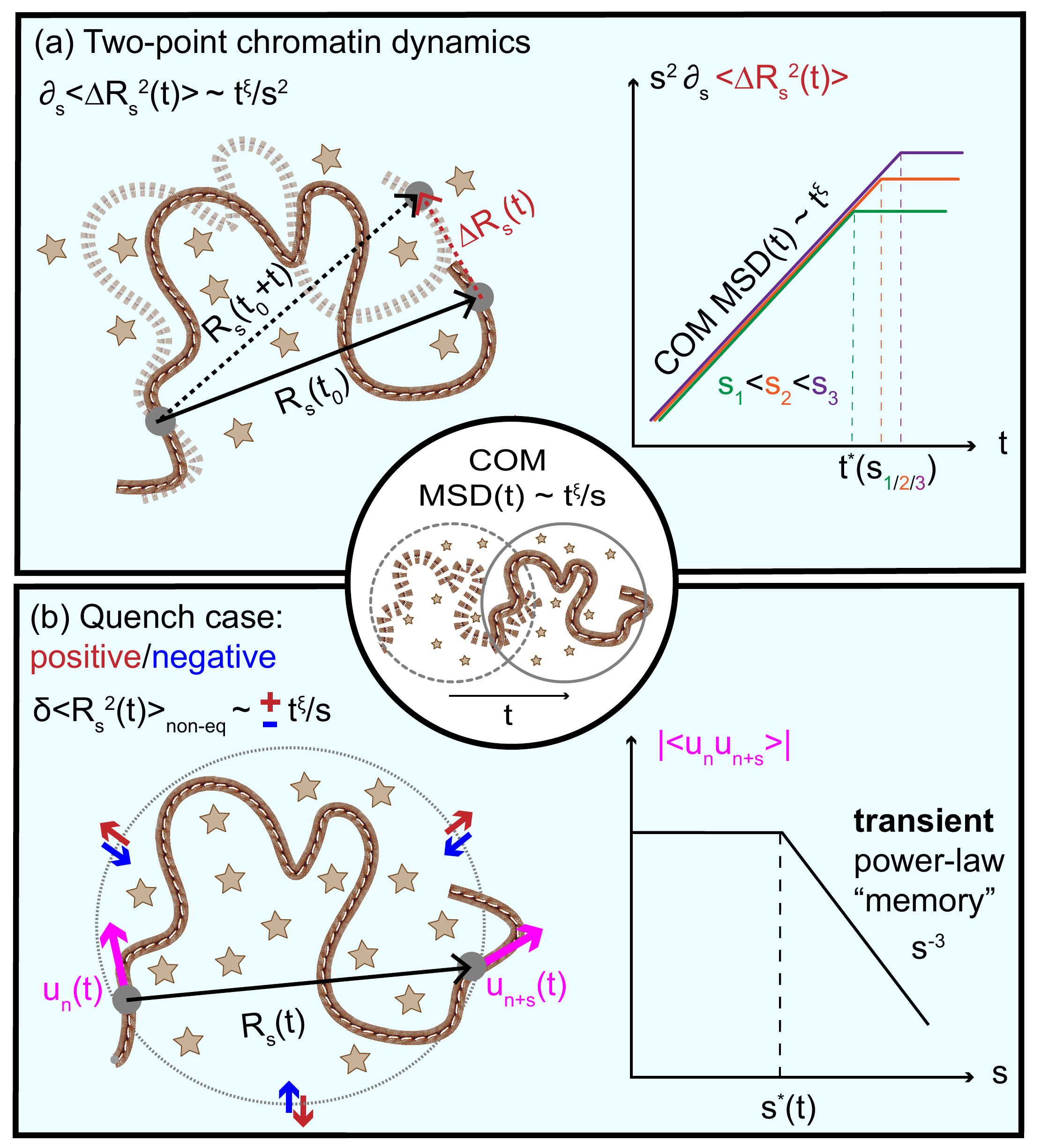}
\caption{Universal center-of-mass (COM) diffusivity $1/s$ for polymer segments of contour length $s$ (center) leads to: (a) corrections to two-point segment dynamics at short times, and (b) transient $s^{-3}$ tangential correlations following a quench, reflecting long-range memory in chain conformation.}
\label{fig:fig3}
\end{figure}

While earlier studies focused on characterizing dynamic exponents; here, we identify \textit{universal} features emerging from collective polymer dynamics in complex and active media. 
Assuming only reciprocal internal forces between beads
(see, however~\cite{goh24}), and spatially uncorrelated stochastic noise, we find that the center-of-mass (COM) diffusivity of a polymer segment exhibits a universal $1/s$ scaling with segment length $s$, independent of structural or environmental details. This COM scaling imprints characteristic dynamic corrections onto two-point fluctuations, which can be observed via recent two-locus chromatin tracking experiments~\cite{gabriele22,bruckner}, enabling access to collective chromatin dynamics based on  two-locus tracking data (Fig.~1a). 

Additionally, we show that following a quench in temperature or activity, center-of-mass dynamics generates transient long-range tangential correlations along the chain, decaying as $\sim 1/s^3$ [Fig.~1b]. This provides a clear signature of non-equilibrium behavior, absent in equilibrium ideal chains. The effect is distinct from the $\sim 1/s^{3/2}$  bond-bond correlations that arise in dense polymer melts due to excluded volume and the correlation hole effect~\cite{wittmer04,wittmer07}.

Our results offer a possible explanation for an apparent mismatch between structural and dynamical measurements of chromatin: Hi-C experiments report fractal globule statistics with $R(s) \sim s^{1/3}$ for genomic distance $s$~\cite{lieberman,mirny11,grosberg93,grosberg88}, while live imaging reveals Rouse-like single-locus MSD $\langle x_n^2(t) \rangle \sim t^{z}$ with $z=1/2$~\cite{bruckner,keizer23,gabriele22}. We suggest this discrepancy reflects a crossover between dynamical regimes: at short times ($t < \tau_e \sim 200$–$500$ s), dynamics appear Rouse-like, while at longer times topological constraints slow motion to $z \approx 0.3$~\cite{smrek15,ge16,halverson14}. Although this slowing is hinted at in single-locus data~\cite{bruckner,keizer23}, limited time windows may obscure the crossover. By analyzing two-locus fluctuations, we extract a correction term with exponent $\xi = 0.77 \pm 0.14$, consistent with theoretical predictions for fractal globules~\cite{smrek15,ge16}. This supports the interpretation of crumpled chromatin and suggests a connection between its structural and dynamical signatures.

\textit{Model:}
We consider a general overdamped Langevin model for the coordinates $\{x_i(t)\}_{i=1}^N$ of polymer beads connected by Hookean springs of rigidity $\kappa$ along the chromatin contour:
\begin{equation}
\begin{split}
\Gamma(3-\alpha) {}_c &D^{\alpha} x_n(t) = \kappa \gamma_\alpha^{-1} \left(x_{n+1}-2x_{n}+x_{n-1}\right)+ \\
&+\gamma_\alpha^{-1}\sum_{k\ne n\pm 1} f^{\text{vol}}_{nk}(\{x_i, \dot{x}_i\}_{i=1}^N, t) + \eta_n\,,
\label{gend}
\end{split}
\end{equation}
where ${}_c D^{\alpha}$ is the Caputo fractional derivative of order $\alpha \le 1$ describing viscoelasticity of nucleoplasm \cite{polovnikovprl18,lampo16,Weber12,Grimm18} and $\gamma_\alpha$ is the corresponding friction coefficient. The term $f^{\text{vol}}_{nk}$ in the r.h.s. of Eq.~\eqref{gend} accounts for non-neighboring (volume) interactions between distant beads of the chain; $\eta_n$ is a Gaussian noise independently acting on different beads with the following covariance:
\begin{equation}
\langle \eta_n(t) \eta_k(t')\rangle = \Theta_1\,\mathcal{C}(t-t') \delta_{nk},
\label{eta}
\end{equation}
where $\mathcal{C}(t-t')$ is the normalized temporal correlation function. The parameter $\Theta_1$ sets the noise strength, and may be proportional to either temperature $T$ in equilibrium or activity $A$ in driven systems.
We note, however, that hydrodynamic effects (as in the Zimm model \cite{doi}) fall outside this framework due to spatially correlated noise.

Chromatin in vivo can be driven by ATP-dependent ``excitations" from motor proteins such as RNA polymerase II or loop-extruding proteins~\cite{Goychuk23} or effectively described as close-to-equilibrium~\cite{polovnikovprl18,polovnikovsm18}. While Eqs.~\eqref{gend}-\eqref{eta} are agnostic to  either case, thermal equilibrium imposes a specific form on the noise correlation function: $C(\Delta t)=(2-\alpha)(1-\alpha)\Delta t^{-\alpha}$. Assuming further that interactions between monomers can be expressed by an arbitrary quadratic form of the bead coordinates~\cite{polovnikovsm18} results in linear dynamics with $f_{nk}^{\text{vol}}\sim x_n-x_k$ converting Eq.~\eqref{gend} into the generalized Rouse-Langevin equation (GRLE, details are given in Supplementary Materials \cite{supp}) that can describe a crumpled chromosome in a viscoelastic fluid at equilibrium~\cite{polovnikovprl18}.


Assuming reciprocity of internal forces ($f^{\text{vol}}_{nk} = -f^{\text{vol}}_{kn}$) in Eq.~\eq{gend}, the stationary dynamics of COM $x_{\text{COM}} = N^{-1}\sum_{i=1}^N x_i$ follows $\Gamma(3-\alpha) {}_c D^{\alpha} x_{\text{COM}}(t) = \eta_{\text{COM}}$. The noise covariance here acquires a $1/N$ factor due to independence of bead--level fluctuations: 
\be
\langle \eta_{\text{COM}}(t)\; \eta_{\text{COM}}(t')\rangle = \Theta_1 N^{-1} \mathcal{C}(t-t'),
\ee
leading to the following universal behaviour of the COM diffusion coefficient:
\be
\Gamma_{\text{COM}}(N) = N^{-1} \Theta_1\,,
\label{dcom}
\ee
where $\Theta_1$ is proportional to the single--bead diffusion constant~\cite{GKh94,Goychuk23}. In particular, reciprocity holds for  GRLE dynamics, where the zeroth normal mode $u_0(t)=\sqrt{N} x_{\text{COM}}$ \cite{supp} yields the well-known $1/N$ behaviour of the COM diffusion coefficient of the generalized Rouse model~\cite{TammPolovnikov,polovnikovprl18}.

Equation~\eqref{dcom} admits a natural interpretation in terms of a scaling picture based on independently relaxing blobs: A polymer segment of contour length $s$ relaxes over a characteristic time $\tau(s) \sim s^{w}.$  At short times $t \ll \tau(s)$  the segment comprises $s/s^*$ independently diffusing blobs of size  $(s^*)^{1/d_f}$ yielding
\be
\langle x_{\text{COM}}^2 (s,t) \rangle \sim (s/s^{*})^{-1} \left(s^{*}(t)\right)^{2/d_f} \sim s^{-1} \;t^\xi,
\label{sc}
\ee
where $\xi = (2+d_f)(w d_f)^{-1}$ is the short--time  MSD exponent of COM. Crucially, the  $s^{-1}$ scaling of  COM diffusivity is independent of the fractal dimension ($d_f$), viscoelastic properties ($\alpha$), and activity level.



\textit{Two-point dynamics:}
The universal COM diffusivity introduces a distinct correction to segment-scale fluctuations. To quantify this, we consider the mean-squared displacement (MSD) of the separation between two monomers a contour distance $s$ apart,
\be
M_2(s, t) = \langle \left(R(s, t) - R(s, 0) \right)^2\rangle\,,
\label{m20}
\ee
where $R(s, t) = x_n(t) - x_{n+s}(t)$ is the segment displacement. This can be written in terms of the auto-correlation $C_2(s,t)=\langle R_s(t) R_s(0) \rangle$ as 
\be
M_2(s, t) = 2C_2(s,0) - 2 C_2(s,t)\,,
\label{m2}
\ee
which vanishes at $t=0$, and saturates to $2 \langle R_s^2 \rangle$ \cite{slavov23} for $t \gg \tau(s) \sim s^{w}$. 
Under  equilibrium GRLE dynamics, $C_2(s,t)$ is fully determined by the mode autocorrelations $C_p(t) = \langle u_p(t) u_p(0) \rangle$ as
\be
C_2(s, t) = 2N^{-1}\sum_{p=1}^N C_p(t) \left(1 - \cos \left(2 \pi p s/N\right)\right).
\label{c2st}
\ee
At short times $t \ll \tau(s),$ this decomposes into
\be
C_2(s,t) = C_2(s,0) - \langle x_n^2(t) \rangle + \langle x_{\text{COM}}^2(s, t) \rangle; \quad t \ll \tau(s),
\label{c21}
\ee
where $\langle x_n^2(t) \rangle = \Theta_1 t^z$ is the individual monomer MSD, and $\langle x_{\text{COM}}^2(s,t) \rangle = B t^\xi s^{-1}$ reflects collective motion. Substituting into Eq.~\eqref{m2} yields gives
\be
M_2(s,t) = 2\langle x_n^2(t) \rangle - 2\langle x_{\text{COM}}^2(s, t) \rangle = 2 \Theta_1 t^z - 2 Bt^\xi s^{-1},
\label{m21}
\ee
where $B \propto \Theta_1 >0$. 
Thus, the short-time two-point fluctuations exhibit a universal negative correction due to COM dynamics, leading to an apparent enhancement of diffusivity at larger genomic separations.

This result can be derived more generally by expressing the segment fluctuations in the COM frame. At short times, the endpoints fluctuate independently around the COM:
$M_2(s,t) = \langle [\Delta R_{cn}(t) - \Delta R_{cm}(t)]^2 \rangle = \langle \Delta R_{cn}^2(t) \rangle + \langle \Delta R_{cm}^2(t) \rangle$ (see Fig.~S1).
Using $\langle x_n^2(t) \rangle = \Theta_1 t^z$ and $\langle x_{\text{COM}}^2(s,t) \rangle = B t^\xi s^{-1}$, Eq.~\eqref{m21} follows, 
enabling direct extraction of the COM dynamics from two-point measurements.
%
Differentiating with respect to $s$, one obtains
\begin{equation}
s^2 \partial_s M_2(s,t) \sim t^\xi; \quad t \ll \tau(s).
\label{m22}
\end{equation}

\textit{Simulations:}
We tested Eqs.~\eqref{m21} and\eqref{m22} using molecular dynamics simulations.  For equilibrated phantom chains, the derivative $\partial_s M_2(s,t)$ grows linearly with time across a range of segment lengths $s$, in contrast to the monomer Rouse scaling $t^{1/2}$ (Fig.~S2e). Rescaling by $s^2$ collapses the curves, confirming the predicted COM correction $-Bt^\xi s^{-1}$  with $\xi=1$ in this ideal case (Fig.~S2e inset). 

We next considered equilibrium dynamics of an unknotted polymer ring in a melt—a well-established model of the fractal globule state relevant to chromatin~\cite{grosberg93,mirny11,polovnikovsm18,polovnikov23prx,polovnikov23pre}: The gyration radius exhibits a crossover from $R_g^2(s) \sim s$ to $s^{2/3}$ (implying $d_f = 3$), while monomer MSDs shift from $\langle x_n^2(t) \rangle \sim t^{1/2}$ to sub-Rouse motion with $z \approx 0.3$, consistent with predictions for the fractal loopy globule ($z = 2/7$)~\cite{ge16} and annealed lattice animal ($z \approx 0.26$)~\cite{smrek15} (Fig.~S3a–b). The same crossover at the entanglement time $\tau_e$ appears in $M_2(s,t)$ (Fig.~S3c):
\be
M_2(s,t) \approx 2\langle x_n^2(t) \rangle \sim 
\begin{cases}
t^{1/2}, \quad t < \tau_e \\
t^{z \approx 0.3}, \quad \tau(s) \gg t>\tau_e
\end{cases}\,.
\label{m2_cg}
\ee
Using $z = 2/7$ and $d_f = 3$, the predicted COM exponent for long segments is $\xi = z(2 + d_f)/2 = 5/7 \approx 0.71$. Simulation results are consistent with this prediction: (1) the rescaled COM fluctuations $s$ $\langle x_{\text{COM}}^2(s,t) \rangle$ collapse onto a $t^\xi$ scaling (Fig.~S3d), and (2) $\partial_s M_2(s,t)$ agrees with the predicted $t^\xi s^{-2}$  dependence (Fig.~S3e). Together, these results confirm that the two-point correction reflects collective segment motion across ideal and fractal polymers.



\begin{figure}
\includegraphics[width=0.5\textwidth]{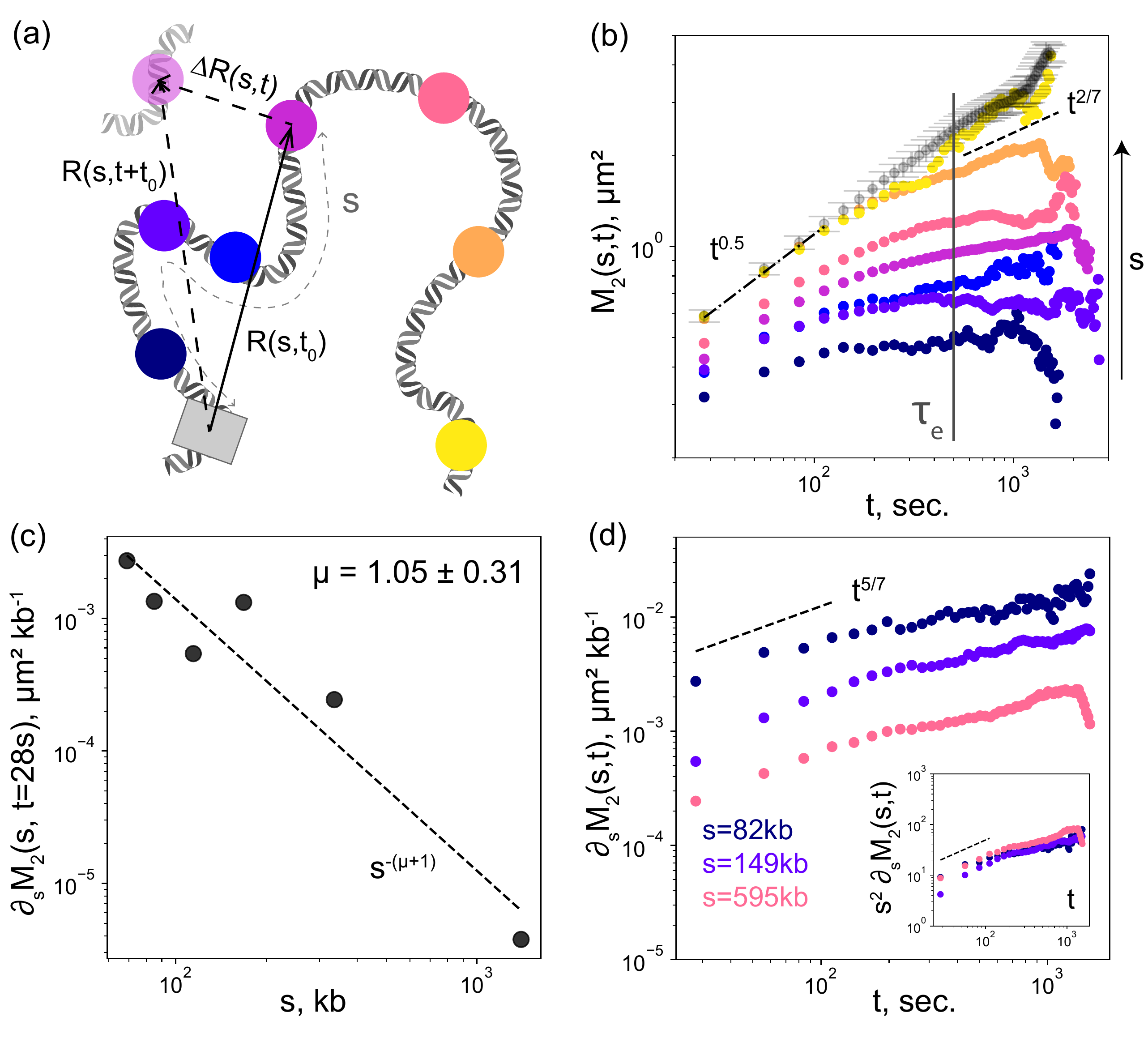}  
        \caption{Analysis of imaging data from~\cite{bruckner}. (a) Experimental design: reporter loci (colored) were placed at genomic distances $s=58,82,88,149,190,595,3300$ kb from a central locus (gray rectangle). (b) MSD of two-point separation $R(s,t)$ for various $s$ (colour corresponds to the panel (a)). The doubled single-locus MSD $2\langle x_n^2(t) \rangle$ is overlaid using gray points, with error bars representing standard deviation across both loci and cells. (c) Derivative  $\partial_s M_2(s, t)$ versus $s$ at initial time point $t=28$s, with best-fit  slope $\mu+1$, where $\mu \approx 1.05 \pm 0.31$. (d) Time evolution of $\partial_s M_2(s, t)$ for three most stable genomic distances $s$. The inset show the curves collapsed by $s^2$ with the exponent of COM MSD for FLG.}
        \label{fig:fig3}  
    \end{figure}

\textit{Analysis of imaging data:} 
Recent advances in two-locus chromatin tracking now allow direct measurement of the two-point MSD $M_2(s,t)$  in live nuclei~\cite{bruckner,gabriele22} (Fig.~\ref{fig:fig3}a). Reanalysis of  data from Ref.~\cite{bruckner} confirms that short-time fluctuations increase systematically with genomic separation $s$ (Fig.~\ref{fig:fig3}b). While earlier interpretations attributed this to scale-dependent two-loci diffusivity, $M_2(s,t) \approx \Theta_2(s) 
 t^z$ \cite{bruckner}, we find that this apparent $s$--dependence instead arises naturally from the COM correction term  $-t^\xi / s$ in Eq.~\eqref{m21}.

A first indication comes from the furthest separations $s=595$ kb and $3.3$ Mb, where $M_2(s,t)$  converges to twice the single-locus MSD, as predicted from Eq.~\ref{m21} for $s\to\infty$. The reduced fluctuations at finite $s$, reflecting collective dynamics, can be quantified by analysis of $\partial_s M_2(s,t)$: At early times (e.g., $t_0 = 28$s), Fig.~\ref{fig:fig3}c reveals power-law scaling $\sim s^{-(\mu+1)}$ with  $\mu = 1.05 \pm 0.31$, in agreement with the expected $s^{-1}$ correction from Eq.~\eqref{m22}. The analysis involving larger time points exhibits similar exponents (see Fig.~S5a-c). These results provide the first experimental support for this universal COM-induced scaling in chromatin dynamics.

The observed Rouse-like behavior in single-locus MSDs, consistent with $z = 1/2$ dynamics~\cite{keizer23,gabriele22,bruckner}, could well reflect transient behavior at times shorter than the entanglement time $\tau_e \sim 200–500 $s (Fig.~\ref{fig:fig3}b), in accord with Eq.~\ref{m2_cg}.
Such crossover  suggests topological constraints emerging only on longer timescales. Using estimates $N_e \sim 10^2$kb and $\tau_0 = (l_k^2 / D_R)^2 \sim 10^{-2}$s, based on Kuhn length $l_k \sim 0.1\, \mu m$~\cite{arbona17} and Rouse coefficient $D_R \approx 5 \times 10^{-2}\, \mu m^2s^{-1/2}$~\cite{bruckner}, we obtain consistent timescales.
The associated entanglement length $N_e$ and time $\tau_e = \tau_0 N_e^2$ can be enhanced in looped, comblike chromosomes~\cite{polovnikov23prx,chan24}. 

Although a slight slowing of single-locus mean-square displacements (MSDs) appears for $t > 200$\,s, the experimental window remains too narrow to fully resolve the crossover (see Fig.~2b and Fig.~S5d). However, the COM motion extracted from two-loci dynamics (Eq.~\eqref{m22}) exhibits no crossover at $\tau_e$, enabling direct extraction of the collective motion exponent $\xi$.
While the original imaging data contains significant noise (particularly in derivatives, see Fig.~S5e), analysis of the most stable curves (Fig.~2d, for $s=82$, 149, and 595\,kb) is consistent with $s^2$ scaling as in Eq.~\eqref{m22}. Remarkably, the resulting $\xi = 0.77 \pm 0.14$ from short-time behavior falls below the Rouse prediction ($\xi=1$) and agrees with crumpled polymer models: $\xi=5/7$ for fractal loopy globule~\cite{ge16} and $\xi \approx 0.65$ for annealed lattice animals~\cite{smrek15} (see \cite{supp}). 
These results provide strongly support the relevance of topological constraints in chromatin dynamics.

\textit{Quench:} 
In living cells, the level of activity (stochastic force) driving chromatin motion can change either through natural processes such as cell cycle progression or induced experimentally through changes in temperature  $T$  (thermal noise) or activity level $A$  in the nucleoplasm (athermal noise).
These changes correspond to abrupt variations in the effective noise strength $\Theta_1$ in Eq.~\eqref{eta}. Motivated by this biological relevance, we use our framework to explore the dynamical signatures of such quenches in polymer systems, focusing in particular on their impact on tangential correlations (Fig.~1b).


The clearest manifestation of a quench is the emergence of transient, long-range \textit{tangent--tangent correlations} along the chain. 
These can be expressed via the segment displacement $R(s,t)$ as 
\begin{equation}
\langle r_n r_m \rangle = \frac{1}{a^2} \left\langle \frac{\partial x_n}{\partial n} \frac{\partial x_m}{\partial m} \right\rangle = \frac{1}{2a^2} \frac{\partial^2}{\partial s^2} \langle R^2(s) \rangle,
\label{corrs}
\end{equation}
At steady state, these correlations are determined solely by the chain’s fractal dimension $d_f$,
\be
\begin{split}
&\langle r_n r_m\rangle\Big|_{\text{eq}} \sim \delta_{s,0} = \delta_{nm}, \quad d_f=2, \\
&\langle r_n r_m\rangle\Big|_{\text{eq}} \sim \frac{(2/d_f)(2/d_f-1)}{s^{2(1-d_f^{-1})}}, \quad d_f \ne 2.
\end{split}
\label{eqtang}
\ee

We now consider a sudden change in noise strength, $\Theta_1^{(1)} \to \Theta_1^{(2)}$ at $t=0$, initiating a transition to a new steady state.  
The new spatial size can be  more ($\Theta_1^{(1)} < \Theta_1^{(2)}$; positive quench) or less ($\Theta_1^{(1)} > \Theta_1^{(2)}$; negative quench) extended, while retaining the original fractal dimension $d_f$ and dynamic exponents.
In the GRLE framework, this quench modifies the normal mode amplitudes, $\phi_p(t) = \langle u_p^2(t)\rangle$, which evolve from $\phi_{p,1}(t=0) = 3T_1 \kappa^{-1} \lambda_p^{-1}$ to $\phi_{p,2}(t \gg \tau_p) = 3T_2 \kappa^{-1} \lambda_p^{-1}$.

At $t \ll \tau(s)$,  the resulting tangential correlations are given by (see \cite{supp} for full derivation):
\be
\langle r_n r_m \rangle\Big|_{\text{non-eq}}  = \frac{1}{Na^2} \sum_{p=1}^{N/s} {\tilde{\phi_p}}(t) \left(\frac{2\pi p}{N}\right)^2 \cos \left(\frac{2\pi p s}{N}\right),
\label{noneqtang}
\ee
where 
$\tilde{\phi}_p(t) \propto 6\Delta T \gamma_\alpha^{-1} t^{\alpha}$ captures the transient response to a quench $\Delta T = T_2-T_1$. Substitution of $\tilde{\phi}_p(t)$ into Eq.~\eqref{noneqtang} yields the key result:
\be
\langle r_n r_m \rangle\Big|_{\text{non-eq}} \sim \frac{\Delta T}{T} \left(\frac{t}{\tau_0}\right)^\xi \frac{1}{s^3}, \quad t \ll \tau(s),
\label{corrs1}
\ee
where $\tau_{0}=\gamma_\alpha \kappa^{-1}$ is the microscopic Rouse time. Equation~\eqref{corrs1} reveals universal $s^{-3}$ scaling of correlations post-quench. The overall sign reflects the direction of the quench, with $\Delta T > 0$ (positive quench) producing positive correlations, and vice versa.

\begin{figure}[t]        
\includegraphics[width=0.5\textwidth]{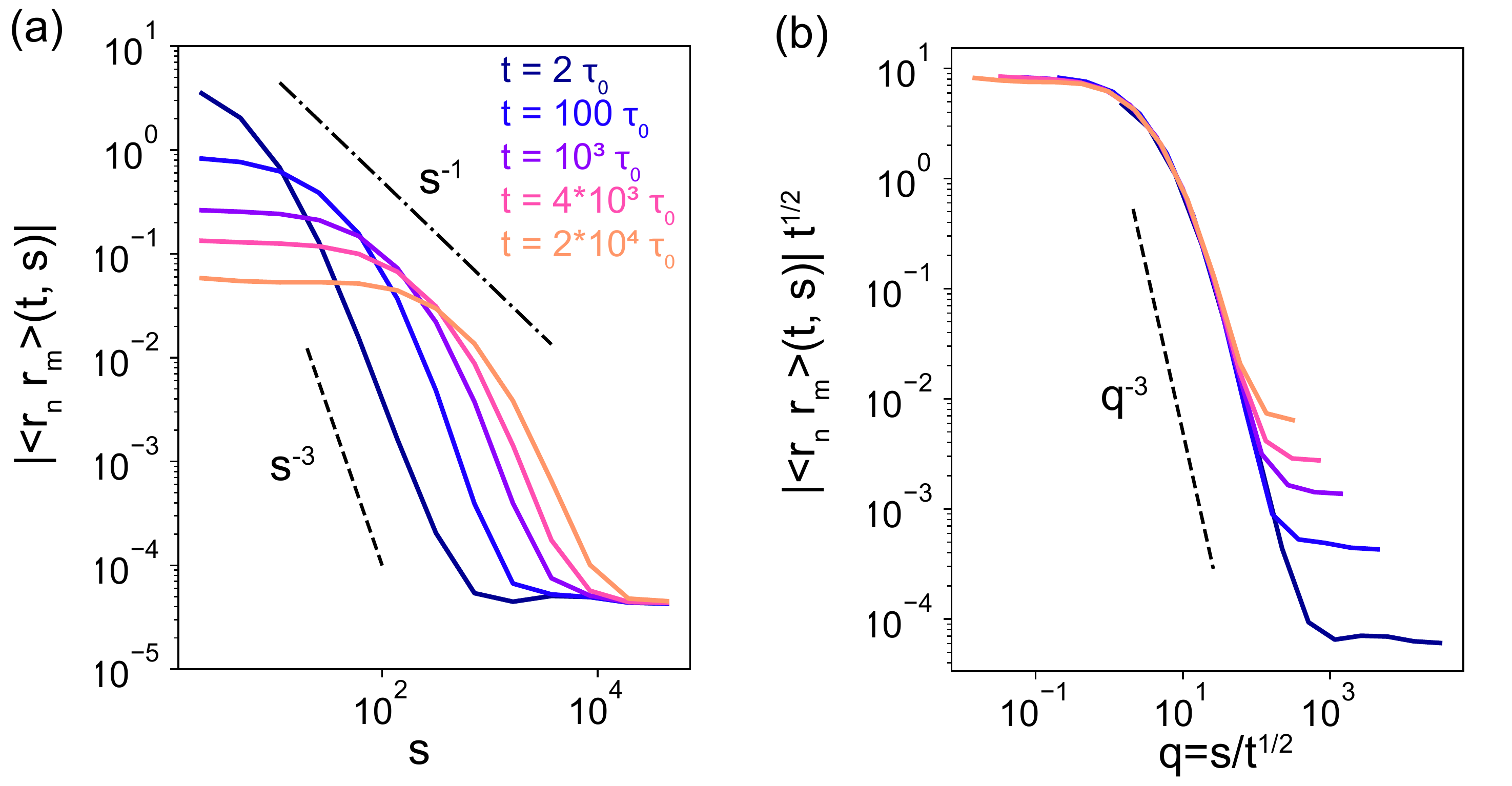}  
        \caption{Absolute value of tangent--tangent correlations $\langle r_n r_m \rangle$ in an ideal chain ($N=10^5$ monomers) following a quench of the microscopic bead spacing from $a=1$ to $a=5$,  obtained from molecular dynamics simulations. (a) Time evolution of  correlations  after the quench. For each $t$,  correlations  decay as $\sim s^{-3}$, for segment lengths $s>s^*=(t/\tau_0)^{1/2}$, and  saturate the  equilibrium level for $s<s^*$. (b) Data collapse of curves from panel (a) under rescaling $s \to s/t^{1/2}$, with normalization applied at $s=2$. The rescaled curves exhibit constant behavior for $s < s^*$  and $s^{-3}$ decay for $s > s^*$.}
        \label{fig:fig2}  
    \end{figure}

At early times, both equilibrium and non-equilibrium contributions are present.  For non-ideal chains ($d_f \neq 2$) the equilibrium contribution decays as $s^{-2(1 - 1/d_f)}$, eventually dominating the universal non-equilibrium term of $s^{-3}$. However, for ideal chains ($d_f=2$) the  equilibrium term vanishes, revealing the pure non-equilibrium signal:
\be
\langle r_n r_m \rangle\Big|_{d_f=2} \sim \frac{1}{s^{*}}\left(\frac{s^{*}}{s}\right)^{3}, \quad s > s^* = (t/\tau_0)^{\xi/2}\,
\label{corrs2}
\ee
that indicates the emergence of transient power-law effects in \textit{ideal} polymer chains.

This prediction was confirmed by simulations of an ideal bead-spring polymer ($N=10^5$) in a viscous fluid. We implemented a quench by instantaneously increasing the bead spacing $a$ from 1 to 5, mimicking an effective temperature increase. As shown in Fig.~2a, for $s > s^*$, the tangential correlations follow the predicted $s^{-3}$ power law, while for $s < s^*$, segments have equilibrated. 
The correlation curves show clear shoulders at $s^*(t)$, confirming the $s^{-1}$ scaling of Eq.~\eq{corrs2}. Rescaling $\langle r_n r_m \rangle \to \langle r_n r_m \rangle t^{1/2}$ and $s \to q = s/t^{1/2}$ collapses the data  (Fig.~2b), with 
constant behavior for $q < 1$ and $ q^{-3}$ decau for $q > 1$.

The physical origin of these transient power-law correlations lies in the COM-driven term in the segment separation dynamics. The separation vector dynamics at $t \ll \tau(s)$ consists of two terms:
\be
\langle \left(x_n(t) - x_{n+s}(t)\right)^2\rangle\Big|_{\text{non-eq}} = A\;t^z +  B\; t^\xi s^{-1},
\label{e1}
\ee
where $A$ and $B$ are some constants. 
The monomer term $A t^z$ dominates the MSD, but does not contribute to the spatial derivative in Eq.~\eqref{corrs}. The $s^{-1}$ COM term instead drives the observed long-range correlations. Thus, the universal $s^{-3}$ scaling emerges as a direct consequence of post-quench collective dynamics.

\textit{Conclusion:} 
We have shown that the assumptions of reciprocal inter-bead forces and spatially uncorrelated noise lead to a universal $1/s$ scaling of the center-of-mass (COM) diffusivity for polymer segments of length $s$. This collective effect imprints measurable corrections on two-point fluctuations, observable across equilibrium, driven, and out-of-equilibrium regimes, with direct implications for chromatin dynamics.
Although non-reciprocal interactions have been proposed in enhancer-promoter communication~\cite{goh24}, the observed universal $1/s$ scaling in the imaging data is consistent with reciprocal dynamics and weakly correlated activity at the segment level. 

Combining an analytically solvable model, scaling arguments, and simulations, we derived a correction term to two-point MSDs arising from COM motion. Reanalysis of  two-locus chromatin tracking data~\cite{bruckner} confirms this prediction and yields a dynamic exponent $\xi\approx 0.77$ consistent with crumpled globule models~\cite{smrek15,ge16}, providing strong evidence for topological constraints in vivo. 
Our results suggest that the Rouse-like behavior reported in~\cite{gabriele22,keizer23,bruckner} reflects transient early-time dynamics of a crumpled polymer, thereby resolving the apparent paradox between chromatin’s fractal structure and Rouse-like motion.


%
%

Finally, we demonstrated that a sudden change in temperature or activity induces transient $s^{-3}$ tangent--tangent correlations—even in ideal chains—driven by universal COM dynamics. These non-equilibrium correlations act as a memory mechanism and extend the phenomenon of broken Flory ideality~\cite{wittmer04,wittmer07} to dynamically perturbed polymer systems.

\textit{Acknowledgements}. The authors are grateful to Andrey Cherstvy, Ralf Metzler and Pietro Luigi Muzzeddu for illuminating discussions. The work of KP is supported by the Russian Science Foundation (Grant No. 25-13-00277). KP also acknowledges the support of the Alexander von Humboldt Foundation. MK is supported by the NSF through grants DMR-2218849 and MCB-2044895.

\bibliography{bibliography}

\end{document}